\documentstyle[12pt]{article}
\def\laq{\raise 0.4ex\hbox{$<$}\kern -0.8em\lower 0.62ex\hbox{$\sim$}}
\def\gaq{\raise 0.4ex\hbox{$>$}\kern -0.7em\lower 0.62ex\hbox{$\sim$}}
\input psfig.tex
\begin{document}
\centerline{\Large\bf Brane-Production and the Neutrino-Nucleon 
}
\centerline{\Large\bf cross section at Ultra High Energies}
\centerline{\Large\bf in Low Scale Gravity Models}

\bigskip
\centerline{\bf Pankaj Jain, Supriya Kar and Sukanta Panda}

\bigskip
\begin{center}
Physics Department\\
I.I.T. Kanpur, India 208016\\
\end{center}
\bigskip

\noindent {\bf Abstract:}
The origin of the ultra high energy cosmic ray (UHECR) showers
has remained as a mystery among particle physicists and astrophysicists.
In low scale gravity models, where the neutrino-nucleon cross section
rises to typical hadronic values at energies above $10^{20}$ eV, the
neutrino
becomes a candidate for the primary that initiates these showers.
We calculate the neutrino-nucleon cross section at ultra high energies
by assuming that it is dominated by the production of p-branes. 
We show, using a generalized Randall-Sundrum model, 
that the neutrino-nucleon cross-section at neutrino energies
of $10^{11}$ GeV is of the order of 100 mb, which is required for explaining
UHECR events. Similar result also follows in other models such as the
Lykken-Randall model.

\newpage
Low scale gravity models \cite{add,RS}
lead to the possibility that the neutrino-nucleon
cross section at ultra high energies may be several orders of magnitude
larger than what is predicted by the Standard Model. At center of
mass energies $\sqrt s$ less than the scale of quantum gravity $M_*$,
which we assume is of order 1 TeV, the cross section can be calculated
using the perturbative Feynman rules \cite{HaLyZh}. However at higher energies,
such that $\sqrt s>M_*$, there does not exist any reliable procedure
to calculate the cross section. In applications to cosmic rays we
are interested in cross section at center of mass energies of the
order of $10^{12}$ GeV$^2$, where the perturbative approach is certainly
not applicable. Several authors
\cite{N+S,us1,d+d,kp,stringy,Emparan2,Feng,Anchordoqui1} have obtained various
estimates by using different models for the cross section in this
energy regime. In Ref.
\cite{us1} we assumed two models for the $s$ dependence of the
cross section, $\sigma\sim s,s^2$, and found that neutrino-nucleon
cross sections are of the order of one to several hundred mb with
the precise value dependent on the model used and on the choice of
$M_*$. The larger value of the cross section is obtained using
the $s^2$ model. Such large values of the neutrino-nucleon cross section
are very interesting since then the neutrino becomes a candidate
\cite{us1,d+d}
for explaining the puzzling cosmic ray events which appear to violate
the GZK bound \cite{GZK,puget,nw,sigl}.
The idea that neutrino-nucleon
cross section grows large at GZK energies is very old \cite{early}.
Large extra dimension, low scale gravity models \cite{add,RS}
have provided new impetus to this idea \cite{us1,d+d,kp,tos,neu,N+S2}.
Even if the cross section is not large enough
to explain the GZK events, cosmic ray observations can be used to
test this hypothesis and to put
stringent bounds on these models.

There exist several proposals for calculating
the ultra high energy scattering cross sections including the
gravitational contribution.
In Ref. \cite{N+S,Emparan2,Emparan1,Rattazzi} the authors proposed an
eikonal model for the cross section, which may be applicable for
very large $s$ but for momentum transfer $-t<M_*^2$. For larger
values of the momentum transfer, Ref. \cite{N+S,Banks,BH1,BH2,Eardley}
have proposed that black hole production will set in.
This proposal has generated great interest with several
studies investigating the production rate of black holes
at colliders \cite{BH2,Hossenfelder,Cheung,Casadio,Ringwald,Park,Rizzo}
and cosmic ray collisions
\cite{Feng,Anchordoqui1,Anchordoqui2,Uehara,Kowalski}.
Using the
eikonal model \cite{Emparan2} and the
black hole production rate \cite{Feng}, 
it is found that the neutrino-nucleon cross section
is of the order of $0.1$ mb at $s\approx 10^{12}$ GeV$^2$.
This cross section
is also large enough such that neutrinos
could generate horizontal showers, which can easily be seen in future
observatories.

Black hole production cross section is estimated by assuming
that it is approximately equal to the geometric area of the black
hole produced. A collision between partons $i$ and $j$ with
center of mass energy $\sqrt{\hat s}$ is assumed to produce a
black hole of mass $M_{\rm BH} \approx \sqrt{\hat s}$ and hence the
parton level cross section $\hat \sigma(ij\rightarrow{\rm BH})$
is equal to $\pi R_S^2$ where $R_S$ is the Schwarzschild radius
of a black hole of mass $M_{\rm BH}$. This geometric cross section
has been criticized by Voloshin \cite{Voloshin} on the grounds
that a high energy collision with partons $i$ and $j$ will
have a very large amplitude to radiate, and the amplitude to produce
an isolated black hole will be exponentially damped. At our current
stage of understanding of quantum gravity, the various models proposed must be
considered as highly speculative.

Recently Ref.  \cite{Ahn2002} has proposed that besides black holes,
high
energy collisions will also produce branes.
They argue that in certain
cases the production cross section of p-branes of mass $M_p\approx\sqrt 
{\hat s}$ is
much higher than that of black holes of the same mass.  The precise value
of the cross section depends on the dimensionality of the brane and the
size of the extra dimensions.  The largest cross section for a p-brane is
obtained when the brane is completely wrapped on the small-size extra
dimensions where it is assumed that there exist $m$ extra dimensions
compactified on length scale of order $L\laq M_*^{-1}$ and the
remaining $n-m$ extra dimensions are compactified on length scale 
of order $L^\prime >> M_*^{-1}$.
The ratio $\Sigma(s;n,m,p\le m)$ of the cross section of the p-brane to black
hole production in this case is given by \cite{Ahn2002}
\begin{equation}
\Sigma(s;n,m,p\le m) \approx \left(L\over L_*\right)^{-2p\over n-p+1}
{\gamma(n,p)^2\over\gamma(n,0)^2}\left(s\over s_*\right)^{w-1\over n+1}
\end{equation}
where $n$ is the total number of extra dimensions, $L_* = M_*^{-1}$,
\begin{equation}
\gamma(n,p) = \left[{8\Gamma\left(n+3-p\over2\right)\over
(2+n)\sqrt{\left(1-{p\over n+2}\right)/(p+1)}}\right]^{w\over n+1}
\end{equation}
and $w=1/[1-p/(n+1)]$.  The branes, once produced, will decay producing 
gravitons and matter fields.

We calculate the neutrino-nucleon total
cross section $\sigma_{\nu p}$
at ultra high energies assuming that the cross section is
dominated by brane production.   
We perform this calculation within the framework of a generalized 
Randall-Sundrum (RS)
model \cite{RS} and the model proposed by Lykken and Randall in Ref. \cite{LR}
such that these models have an arbitrary number of compact extra dimensions
instead of the single compact dimension assumed in Ref. \cite{RS,LR}.
Besides the single compact dimension which involves the warp factor,
the remaining $m$ compact dimensions are assumed to be factorizable. 
These $m$ extra dimensions are compactified on a m-torus of radius R.
The generalized RS model, therefore, 
consists of two 3-branes with opposite tension
situated at orbifold points $y=0,y_c$ in
a (4+m+1)-dimensional space-time. 
The metric can be written as
\begin{equation}
ds^2_{D+1} = e^{-2\sigma(y)}\eta_{\mu\nu}dx^\mu dx^\nu 
+ R^2\delta_{ab}d\theta_a d\theta_b + dy^2 
\end{equation}
where $\mu,\nu=0,..,3$, $a,b=1,...,m$,  
$\theta_a$ are the compactied dimension with range [0,2$\pi$]
and y is the warped dimension with range [-$y_c,y_c$].
The Einstein equations can be written as
\begin{equation}
G_{AB}=\frac{1}{4 M^{m+3}}\left(T_{AB}^{(B)}+T_{AB}^{(br)}\right)
\end{equation}
The components of the bulk energy-momentum tensor are assumed to be
$$ T^{(B)\mu}_\nu  = \delta_\nu^\mu \Lambda_0,~~ T^{(B)y}_y =
\Lambda_0,
~~T^{(B)b}_a = \Lambda_\theta\delta_a^b $$
The components of the brane energy-momentum tensor are assumed to be
$$ T^{(br)\mu,i}_\nu  = \delta(y-y_i) v_0^i \delta_\nu^\mu ,~~
T^{(br)y,i}_y
=0, 
~~T^{(br)n,i}_m =  \delta(y-y_i) v_\theta^i  \delta_m^n\ , $$
where i=1,2,
and all other non-diagonal components are taken to be zero.
Inhomogeneities in the energy-momentum tensors and cosmological
constants can be due to different
contributions to casimir energies in the different direction of
space-time \cite{kv,cw}.
In the absence of a four dimensional cosmological constant, the
($\mu,\nu$)-component of the Einstein's equation gives
\begin{equation}
-3 \sigma''(y)+6\sigma'^2(y)=-\frac{\Lambda_0}{4
M^{m+3}}-\sum_i\frac{v_0^i}{4
M^{m+3}}\delta(y-y_i)
\label{ein1}
\end{equation}
The (y,y)-component of the Einstein's equation is given by
\begin{equation}
6 \sigma'(y)^2 = -\frac{\Lambda_0}{4 M^{m+3}}
\label{ein2}
\end{equation}
The ($\theta,\theta$)-component of the Einstein's equation is given by                
\begin{equation}                                         
 -4\sigma''(y) +10 \sigma'^2(y) = -\frac{\Lambda_\theta}{4 M^{m+3}}
-\sum_i\frac{v_\theta^i}{4 M^{m+3}}\delta(y-y_i)
\label{ein3}
\end{equation}
From Eq.(\ref{ein1},\ref{ein2},\ref{ein3}), using the orbifold symmetry
along y-direction, we find
\begin{equation}
\sigma(y)=\sqrt{-\frac{\Lambda_0}{24 M^{m+3}}}~(|y|-y_c)
\end{equation}
and $\Lambda_\theta=\frac{5}{3}\Lambda_0$,~~$v_\theta^i=\frac{4}{3}v_0^i.$
This solution is valid only if brane tensions and cosmological constants are
related as follows 
\begin{equation}
\Lambda_0 = -24 M^{m+3}k^2,~~v_0^1=-v_0^2=24 M^{m+3}k
\end{equation}
The four-dimensional reduced Planck mass $M_{pl}$ can be written as
\begin{equation}
M_{pl}^2 =  M^{(m+3)}\exp{(2 k y_c)}{(2\pi R)^m\over k }
\left[1-\exp(-2 k y_c))\right]
\end{equation}
where $M=\frac{M_*}{(32\pi)^{1/(m+3)}}$. If we demand $k y_c \simeq 35,
M_*=1TeV$ and $R<(TeV)^{-1}$
we find that $k<M_*$. We can take $y_c$ of order fm without violating the
existing observational constraints from astrophysics and
accelerator physics and then we find $k$ is of order GeV. We
assume that some unknown mechanism stabilizes the radius of extra
dimensions.

We next determine the spectrum of the KK-modes of this theory in
the low energy limit.
Decomposition of KK states can be carried out by considering the
following
graviton  perturbations:
\bigskip
\begin{equation}
ds^2 = e^{-2 \sigma}[ \eta_{\mu\nu} + h_{\mu\nu}(x,y,\theta_i)]
+ dy^2 + R^2 \delta_{ij} d \theta_i d \theta_j
\end{equation}
where R is the radius of compact extra dimensions and $ \sigma=k
(|y|-y_c)$. Here
we are not considering  fluctuations around extra dimensions.
We expand the linear fluctuations around the flat 4-dimensional metric in
a
complete
set of radial wavefunctions and fourier modes:
\bigskip
\begin{equation}
h_{\mu\nu}(x,y,\theta_i) =
\sum_{n,l}h_{\mu\nu}^{(n,l)}(x)\phi_{(n,l)}(y)
e^{il\theta}  
\end{equation}
Working in a gauge $\eta^{\alpha
\beta}\partial_{\alpha}h_{\beta\gamma}^{(n,l)}=\eta^{\alpha
\beta}h_{\alpha\beta}^{(n,l)}=0 $ and
$\partial^2 h_{\mu\nu}^{(n,l)}(x)=m_n^2  h_{\mu\nu}^{(n,l)}(x),$
the differential equation for radial part is given by
\begin{equation}
\partial_y \left( e^{-4 \sigma} \partial_y \phi_{n,l}(y)\right)=
- e^{- 2 \sigma}\left( m_n^2-\frac{l^2}{R^2}e^{- 2 \sigma}\right)
\phi_{n,l}(y)
\label{radial}
\end{equation}
with normalization condition $(2\pi R)^m\int_{- y_c}^{y_c} dy
e^{-2 \sigma} \phi_n^* \phi_{n'}=\delta_{nn'}$.
After changing variables to $z_n=\frac{M_n}{k}e^{\sigma}$  
and $f_{n,l}=e^{- 2 \sigma}\phi_{n,l}$, Eq.(\ref{radial}) can be
written as (for $y\ne0,y_c$)
\begin{equation}
z_n^2 \frac{d^2f_{n,l}}{dz_n^2}+z_n\frac{d
f_{n,l}}{dz_n}+\left[z_n^2-\left(4+\frac{l^2}{ k^2 R^2} 
\right)\right]f_{n,l}=0
\end{equation}
In the low energy limit we can set
l=0 since these modes will have energy larger than $1/R$.
Then in this limit the theory reduces to RS model except for the fact
that the size of the warped extra dimension is taken to be of order 1 fm
which is about three orders of magnitude larger than the fundamental length
scale of $(1 TeV)^{-1}$. 
The masses of these KK-modes are quantized in mass scale of the order of GeV and
their couplings to the standard model
fields which lives on the brane at $y=y_c$ are the same as in RS case. 
As there are a large number of modes present, we can ignore the few low 
lying modes and the sum over the remaining modes
can be done by changing the summation to
an integration as in the ADD case. 

We also consider another model proposed by Lykken and Randall \cite{LR}
in which the size of the extra dimension can be infinite. The model
was originally proposed with only one extra dimension. The model
involves three branes, one located at $y=0$, second at $y=\infty$ and
the third, which is the physical brane, located at some position
$y=y_0$ such that $0<y_0<\infty$. We generalize
this model by introducing $m$ additional compactified dimensions in
exact analogy to the above construction in the case of RS model. 
We find that the low energy predictions of the model remain unchanged due
to the addition of $m$ compact extra dimensions. Hence we find that
we can extend both the RS and the Lykken-Randall model such that they 
contain $m$ extra dimensions compactified on a length scale of the order
of inverse TeV and one large extra dimension of length scale larger
than (or order of) 1 fm. We can now construct p-brane solutions in these
models such that the brane is completely wrapped around the $m$ extra
compact dimensions. 

The parton level brane production 
cross section $\hat\sigma_{\rm brane}$ for p-brane
of mass $M_p=\sqrt {\hat s}$
is given by,
\begin{equation}
\hat\sigma_{\rm brane}(\hat s) = \Sigma(\hat s;n,m,p\le m) \hat\sigma_{\rm BH}
(\hat s)
\end{equation}
where $\hat\sigma_{\rm BH}$ is the production cross section for
black hole of mass $M_{BH}=\sqrt {\hat s}$. The black hole production 
cross section,
$\hat\sigma_{\rm BH}$, is given by \cite{BH1,BH2}
\begin{equation}
\hat\sigma_{\rm BH} \approx \pi r_S^2
\label{BlackDisc}
\end{equation}
where $r_S$ is the Schwarzchild radius of a $4+n$ dimensional black hole,
\begin{equation}
r_S = {1\over \sqrt \pi M_*}\left[M_{\rm BH}\over M_*\right]^{1\over 1+n}
\left[{8\Gamma\left({3+n\over 2}\right)\over 2+n}\right]^{1\over 1+n}
\end{equation}
The black hole and the brane
production processes are expected to give dominant contributions
when $s>> M_*^2$.

The total cross section can be computed using
\begin{equation}
\sigma(\nu N\rightarrow {\rm brane}) = \sum_i\int_{x_{\rm min}}^1 dx
\hat \sigma_{{\rm brane},i}(xs)f_i(x,Q)\ ,
\end{equation}
where $f_i$ is the parton distribution function corresponding to
the $i^{th}$ parton, $x_{\rm min} = M_*^2/s$
and $Q$ is the typical momentum scale of the collision
which is taken to be the brane mass. We use the CTEQ parton distributions
\cite{CTEQ}
for our calculation. Since the parton distributions are not known
beyond $Q=10$ TeV, we set $Q$ equal to this value if the brane
mass exceeds 10 TeV. The minimum value of $x$ is obtained by assuming
that branes are produced with masses $M_p$ greater than $M_*$.

\begin{figure}[t,b]
\caption{The neutrino-nucleon cross section $\sigma_{\nu p}$
including the p-brane production for several representative
values of $p$ and the number of extra dimensions $n$. The ratio
of the length scales $L/L_*$ of the small to large extra dimensions
has been set equal to 0.25 and the scale of quantum
gravity $M_*=1$ TeV. The highest energy HERA data point
and the cross section obtained in Standard Model (SM) are also shown
for comparison.}
\bigskip
\hbox{\hspace{0em}
\hbox{\psfig{figure=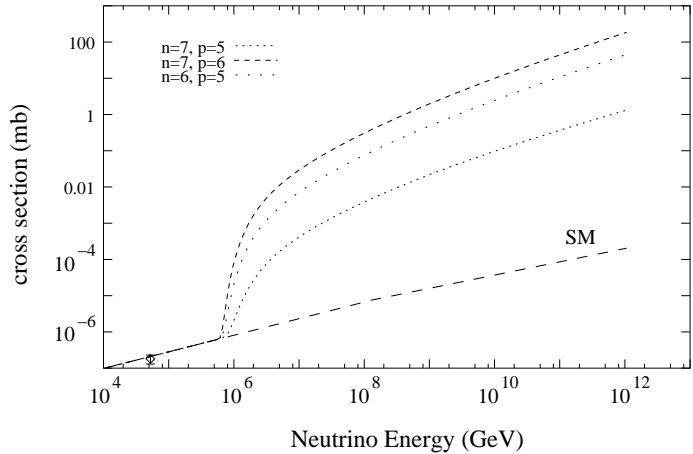,height=6cm}}}
\label{fig_sigma1}
\end{figure}

\begin{figure}[t,b]
\caption{
The $L/L_*$ dependence of the
neutrino-nucleon cross section $\sigma_{\nu p}$
  for several representative
value $p$ and the number of extra dimensions $n$.
he scale of quantum
gravity $M_*=1$ TeV and the incident neutrino energy is taken to be
$10^{11}$ GeV.
}
\bigskip
\hbox{\hspace{0em}
\hbox{\psfig{figure=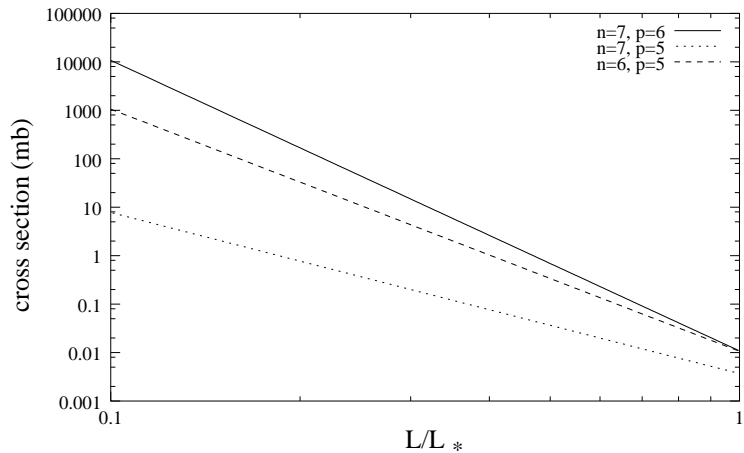,height=6cm}}}
\label{fig_sigma2}
\end{figure}

In fig. \ref{fig_sigma1} we plot the neutrino-nucleon cross section
for some representative choice of parameters, $p$ and $n$. The
ratio of length scales $L/L_*$ has been set equal to 0.25. We find that
for $n=7$ and $p=6$ the cross section rises to roughly 50 mb at
neutrino energy of $10^{11}$ GeV. As $p$ becomes smaller the cross section
is much smaller. Similarly the cross section is considerably reduced
as we lower $n$. The cross section is also found to be very sensitive
to the precise value of $L/L_*$. The dependence of total cross section
on $L/L_*$ is shown in fig. \ref{fig_sigma2} for some representative
choice of parameters, $p$ and $n$. The neutrino energy has been
taken to be $10^{11}$ GeV for this plot.

It is clear from fig.  \ref{fig_sigma1} and fig.  \ref{fig_sigma2} that
there exists a large range of parameter space where the neutrino-nucleon
cross section is of the order of 100 mb or above for the primary neutrino
energy of $10^{11}$ GeV. These are comparable to the nucleon-nucleon cross
section at these energies and is typically the cross section values
required in order that the neutrino can considered as a candidate for
generating the observed air showers above the GZK bound.  We point out that
in the present case the neutrino delivers its entire energy in a single
collision.  This is in contrast to our earlier work \cite{us2}, 
using a t-channel
exchange in which the neutrino loses only a fraction of its energy in a
single collision.  In that case, the neutrino collides with several air
nuclei along its path, and the resulting air shower tends to be longer in
developing compared to an equally-energetic proton shower.  Nevertheless,
the characteristics of air showers do not rule out this model, given
fluctuations and uncertainties in energy \cite{us2}.  In the present case
with high inelasticity, the position of the shower maximum will be
determined dominantly by the value of the neutrino-air cross section.
Since there exists a large range of parameter space where the
neutrino-nucleon cross section exceeds 100 mb, the shower maximum could
exist at the same or higher altitude than that produced by a proton or even
an iron nucleus of equivalent energy.  Hence the position of the shower
maximum may be indistinguishable from a hadron primary of
equivalent energy.

We point out that our cross section does not violate any bounds
imposed by unitarity. The models and parameter ranges 
considered in this paper are not ruled
out by the low energy data. The cross section estimate is based on
the calculation of the area of the classical brane solution, which essentially
can be regarded as an estimate of the size of the region where the
gravitational interaction between the two particles is strong. Therefore
within these models, the total cross section has to be atleast as large 
as the
estimate obtained by assuming that it is dominated by p-brane production.
 Furthermore  as argued in Ref. \cite{us2} it is not possible
to determine whether a cross section violates unitarity simply
by looking at its rate of growth with energy. For example, the large 
$\sigma_{\nu-p}$ 
($>100$mb) obtained in Ref. \cite{us2} by assuming that it rises as 
$s^2$ with energy, 
is well within the bounds 
obtained by Goldberg and Weiler \cite{gw} by a model independent 
application of dispersion relations and unitarity. The growth
of p-brane production cross section with energy is relatively mild
in any case, with the total cross section rising at most as $E_\nu^{0.6}$ at
large energy
for the parameter ranges considered in this paper. 

After formation the p-brane will decay. In analogy with black hole
evaporation, we expect that branes will decay into gravitons and
matter fields. The emitted gravitons will not be detected.
Hence only a
fraction of the particles that are produced will be responsible for
generating the shower and the primary energy may be significantly
underestimated. This implies that for a shower of energy $E$ we need a primary
neutrino of energy  $E'$ which is somewhat larger than $E$.
The precise amount of energy which goes into gravitons is model
dependent.

In conclusion we have shown that in low scale gravity models the
neutrino-nucleon cross section is of the order of or larger than 100 mb for
a large range of parameter space allowed by current experimental
constraints.  This makes the neutrino a candidate for explaining the cosmic
ray events which appear to violate the GZK bound.  Future cosmic ray
observatories will be able to confirm or rule out this hypothesis.  In
particular, the ability of neutrino events to point back to their sources,
which can be established by well-defined statistical
correlations \cite{pointing}, has the potential to establish or to rule out
neutrinos as primaries in the GZK-violation mystery. A very clear discussion
on the possible origin of the highest energy cosmic rays is
given in Ref. \cite{ElbertandSommers}.

\bigskip
\noindent
{\bf Acknowledgements:}
We thank Douglas McKay and John Ralston for useful discussions. 
This work was supported in part by 
a grant from DST (India).

\end{document}